\documentclass{aa}
\usepackage{graphicx}
\usepackage{times,amssymb,lscape,verbatim}
\usepackage{natbib}
\bibpunct{(}{)}{;}{a}{}{,} 
\usepackage[section]{placeins}
\begin{document}

\title{Peculiar spectral and power spectral behaviour\\ of the LMXB GX 13+1}

\author{R. S. Schnerr\inst{1}
	\and T. Reerink\inst{1}
	\and M. van der Klis\inst{1}
	\and J. Homan\inst{1,2}
	\and M. M\'endez\inst{1,3}
	\and R. P. Fender\inst{1}
	\and E. Kuulkers\inst{4}}
\offprints{R. S. Schnerr, \email{rschnerr@science.uva.nl}}
\institute{Astronomical Institute ``Anton Pannekoek'',
	University of Amsterdam, and Center for High Energy Astrophysics,
	Kruislaan 403,
	1098 SJ Amsterdam,
	The Netherlands
	\and INAF - Osservatorio Astronomico di Brera,
	via Bianchi 46,
	23807 Merate (LC),
	Italy
	\and SRON,
	National Institute for Space Research,
	Sorbonnelaan 2,
	3584 CA Utrecht,
	The Netherlands
	\and ESA-ESTEC,
	Science Operations \&\ Data Systems Division,
	SCI-SDG,
	Keplerlaan 1,
	2201 AZ Noordwijk,
	The Netherlands
}

\date{Received 10 Dec 2002 / Accepted 06 May 2003}

\abstract{
We present results of an analysis of all 480 ks of Rossi X-ray Timing Explorer Proportional Counter Array data obtained from 17 May 1998 to 11 October 1998 on the luminous low mass X-ray binary GX 13+1. We analysed the spectral properties in colour-colour diagrams (CDs) and hardness-intensity diagrams (HIDs) and fitted the power spectra with a multi-Lorentzian model. GX 13+1 traces out a curved track in the CDs on a time scale of hours, which is very reminiscent of a standard atoll track containing an island, and lower and upper banana branch. However, both count rate and power spectral properties vary along this track in a very unusual way, not seen in any other atoll or Z source. The count rate, which varied by a factor of $\sim$1.6, along a given track first decreases and then increases, causing the motion through the HIDs to be in the {\em opposite} sense to that in the CD, contrary to all other Z and atoll sources. Along a CD track, the very low frequency noise uniquely {\em decreases} in amplitude from $\sim$5 to $\sim$2\% (rms). The high frequency noise amplitude decreases from $\sim$4\% to less than 1\% and its characteristic frequency decreases from $\sim$10 to $\sim$5 Hz. The 57-69 Hz quasi-periodic oscillation (QPO) found earlier is also detected, and no kHz QPOs are found.
In addition the entire track shows secular motion on a time scale of about a week. The average count rate as well as the amplitude of the very low frequency noise correlate with this secular motion. We discuss a possible explanation for the peculiar properties of GX 13+1 in terms of an unusual orientation or strength of a relativistic jet.
}

\authorrunning{R. S. Schnerr et al.}
\titlerunning{Peculiar behaviour of the LMXB GX 13+1}
\maketitle

\section{Introduction}
\label{intro}
\begin{figure*}[htbp]
\includegraphics[width=0.7\hsize]{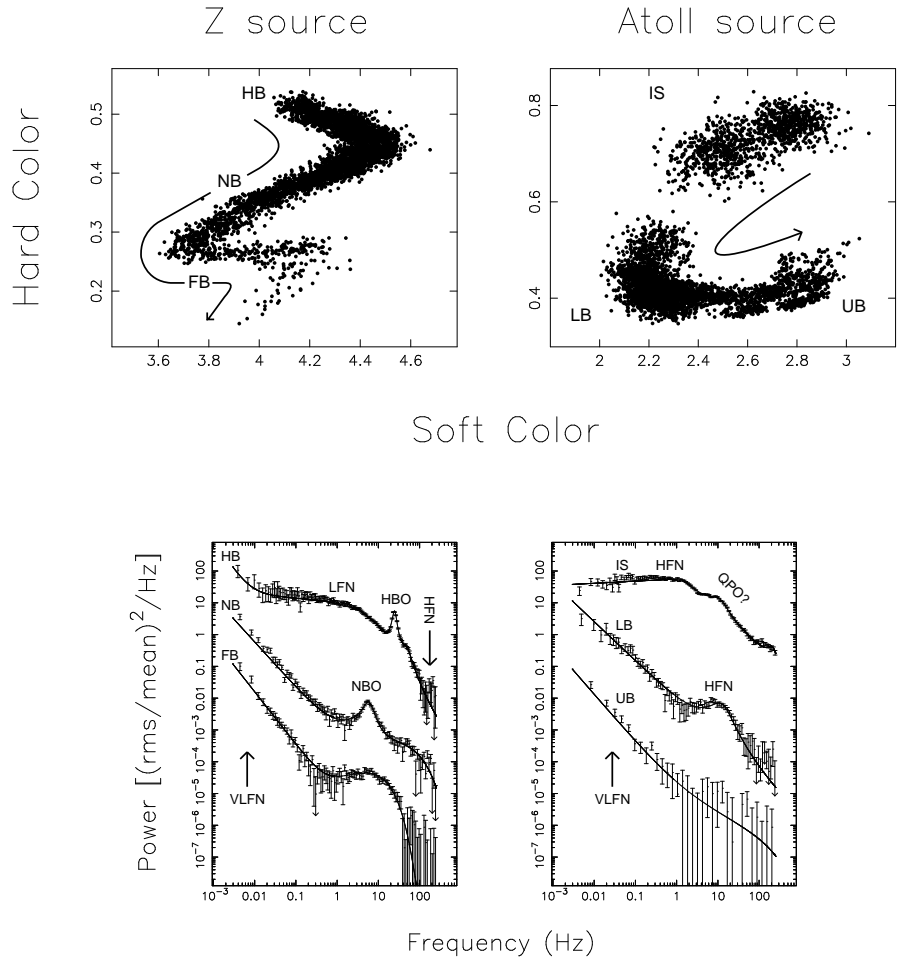}
\caption[]{Typical CDs (top) of Z (left) and atoll (right) sources, and power spectra (bottom) corresponding to their different branches in the CD. The power spectra of the HB and the island state (IS) were shifted upwards by a factor of $10^4$ and those of the NB and lower banana (LB) by a factor of $10^2$. The figure was taken from \citet{rudy:2001}.}
\label{z_atoll}
\end{figure*}
Low mass X-ray binaries (LMXBs) containing a neutron star can be divided into two subclasses, atoll and Z sources, based on their correlated X-ray timing and spectral properties \citep{michiel:1989}. In the colour-colour diagram (CD) LMXBs tend to trace out a well-defined, fully connected 1-dimensional track on a time scale of hours to days for Z sources and days to weeks for atoll sources.

Z sources trace out a Z-shaped track in the CD, where the three branches of the Z from top left to bottom right are called the horizontal branch (HB), the normal branch (NB) and the flaring branch (FB). The (sharp) turns in the track separating these branches are called vertices. 
The power spectral features found in Z sources are very low frequency noise (VLFN), which has an approximately power law shape and is observed up to a few Hz, a band limited noise (BLN) component with a characteristic frequency of 2 to 20 Hz and various different quasi periodic oscillations (QPOs) with a Lorentzian shape and frequencies that range from a few Hz to more than a kHz \citep{michiel:1995,michiel:2000}.

These power spectral features occur correlated to the position of the source in the Z track (see Fig. \ref{z_atoll}). QPOs with frequencies of 13-55 Hz are found on the HB (HBOs), and with frequencies of 4-7 Hz on the NB (NBOs); in some sources the latter continue onto the FB where the frequency increases up to $\sim$20 Hz.
BLN with a fractional root-mean-square amplitude (rms) of up to $\sim$8\% is found along the HB and disappears as the source moves onto the NB. There is a strong tendency of the amplitude of the VLFN to increase from less then 1\% to $\sim$10\% along the whole track from the HB to the FB. The flux increases from left to right along the HB, stays approximately constant or slightly decreases along the NB, and usually increases again on the FB. The mass accretion rate ($\dot{M}$) is thought to increase along the HB via the NB to the FB on time scales of hours to days \citep{michiel:1989}, although on longer time scales the relation with $\dot{M}$ may be more complicated \citep[e.g. ][]{kuulkers:1994,jeroen17+2:2002}.

Atoll sources trace out a track that generally has a C or U-shape, consisting of an island branch, which often takes the shape of one or more isolated patches because of observational windowing (usually in the upper left region of the CD), and a banana branch (extending from the lower banana (LB) --lower left-- to the upper banana (UB) --upper right). VLFN, BLN and QPOs similar to those in Z sources also occur in atoll sources \citep[see Fig. \ref{z_atoll}; ][]{rudy:2001}. Along an atoll-source track from the island to the lower, middle and then to the upper banana, the rms amplitude of the VLFN increases from $\sim$2 to 5\%,  the characteristic frequency of the BLN, which sometimes consists of several components \citep{steve:2002}, first increases and then decreases within a range of $\sim$2 to $\sim$70 Hz, and its rms decreases from $\sim$20\% to less than 1\%. The count rate increases by a factor of $\sim$2-10 \citep{michiel:1989} going from the island state to the banana state (except for the transient sources, see below). The frequencies of the various QPOs increase in the same sense along the track in Z and atoll sources.

GX 13+1 is a LMXB containing a neutron star, since X-ray bursts have been found from this source \citep{fleischman:1985,matsuba:1995}. Together with GX 3+1, GX 9+1 and GX 9+9, GX 13+1 forms the subclass of persistently bright atoll-sources (the ``GX atoll sources'') which were observed to be in the banana state by \citet{michiel:1989}. In GX 13+1 and GX 3+1, two branched structures have been observed in the CD and hardness-intensity diagram (HID) \citep{stella:1985,lewin:1987,schulz:1989,jeroen:1998,muno:2002}. In luminosity, these sources are intermediate between the very luminous Z sources and the weaker remaining atoll sources \citep{ford:2000}. Contrary to the other atoll sources and the Z sources, these GX atoll sources have so far not shown kHz QPOs. Although GX 13+1 has been classified as an atoll source based on upper-banana branch phenomenology \citep{michiel:1989}, among the atoll sources it is the source that shows X-ray properties closest to those of the Z sources \citep[VLFN amplitude: ][ CD track: Muno et al. 2002, $\sim$65 Hz QPOs: Homan et al. 1998]{michiel:1989}. The radio luminosity of GX 13+1 is similar to that of the Z sources, whereas the other atoll sources are weaker \citep{garcia:1988,fender:2000}. From infra-red spectroscopy \citet{bandy:1999} conclude GX 13+1 might be a Z source, based on the nature of its companion suggesting a long orbital period.

High resolution spectroscopic observations of GX 13+1 with ASCA \citep{ueda:2001} and XMM-Newton \citep{sidoli:2002}, have revealed the presence of narrow absorption lines, similar to those in the strong galactic jet sources GRS 1915+105 and GRO J1655-40 and the dippers MXB 1659-29 and X 1624-49.

For some Z sources it is known that the position of the track they trace out in the CD drifts on a time scale of weeks, with shifts in the colours of up to $\sim$10\% \citep{kuulkers:1994,Kuulkers:1996}; this effect is called secular motion. It is also seen in the atoll source 4U 1636-53 \citep{prins:1997,tiziana:2002}, but in the CD it is not as strong as in some Z-sources. In GX 13+1, \citet{muno:2002} also detected secular motion \citep[see also ][]{jeroen:1998}. These shifts are usually larger in the HIDs compared to the CDs due to the shift of the track being strongly correlated with X-ray luminosity ($L_\mathrm{x}$). The power spectral properties of all LMXBs seem to depend primarily on the position \citep[$S$, ][]{Hasinger:1990} of the source in the atoll or Z track and not on $L_\mathrm{x}$. $L_\mathrm{x}$ shows a strong correlation (usually positive, sometimes negative) with this $S$ on short time scales (hours to days), but this correlation can be completely absent on long time scales. It has long been assumed that this is because $S$ is a measure of the accretion rate $\dot{M}$, with $L_\mathrm{x}$ not proportional to $\dot{M}$ due to uncertain causes such as, e.g., beaming, but more recently a scenario has been suggested in which the inner radius of the accretion disc ($r_\mathrm{inner}$) determines the power spectral properties, while at the same time $L_\mathrm{x}$ is determined by the total mass accretion plus perhaps nuclear burning \citep{michiel:2001}. The decorrelation of $L_\mathrm{x}$ from the power spectral properties occurs in this scenario through energy release that does not originate from accretion through the inner disc, but does respond to inner disc count rate changes in a time-averaged sense.

Recently, \citet{muno:2002} and \citet{gierlinski:2002} suggested that the distinction between atoll and Z sources might be an artifact of incomplete sampling, because the atoll sources that cover a wide enough range in count rate \citep[the transient sources Aql X-1, 4U 1608-52 and 4U 1705-44, each with a flux range of more than a factor of 10, see also][]{barret:2002} also exhibit roughly Z-shaped tracks with at the lowest luminosities a 'horizontal branch'. However, the differences in flux ranges and time scales on which these tracks are traced out in the CDs  \citep[as also noted by ][]{muno:2002} as well as their topology and power spectral behaviour \citep{steve:epsilon}, argue against any simple unification of these two classes. Being a persistently bright source with a flux range of a factor of $\sim$2, GX 13+1 is not among the sources that has been observed to show such an atoll source 'horizontal branch'.

In this paper we discuss the X-ray colour, flux and timing properties of GX 13+1 and their mutual dependence. We find highly unusual power spectral -- colour-colour diagram behaviour, which does not fit in with either Z or atoll behaviour as well as very unusual secular motion. In Sect. \ref{explanation} we suggest a possible explanation for the unusual properties we find in GX 13+1.

\section{Observations}
We used all publicly available Rossi X-ray Timing Explorer \citep[RXTE,][]{bradt:1993} Proportional Counter Array \citep[PCA,][]{jahoda:1996} data of GX 13+1 in the third RXTE gain epoch, a total of 44 observations (listed in Table \ref{ptracks}) ranging from May 17 to October 10, 1998 and lasting about 480 ksec in total. Individual observations typically last $\sim$10 ksec, interrupted several times by South Atlantic Anomaly passages and Earth occultations. The uninterrupted parts of observations are normally $\sim$3 ksec in duration. In 33 observations the five Proportional Counter Units (PCUs) 0--4 were all active, but in 11 observations PCU 3 and/or 4 were inactive for part or for the entire observation. The 2-60 keV count rate for 5 PCUs varied between $\sim$3300--5500 cnts $\mathrm{s^{-1}}$. For all observations Standard 2 data are available which have a time resolution of 16 s in 129 energy channels (covering the 2-60 keV PCA range), as well as high time resolution Single Bit and Event data with a time resolution of 1/4096 s or better, also covering the 2-60 keV PCA range.

Standard 2 data of PCUs 0, 1 and 2 (which were always active) were used to create CDs and HIDs with 16 s data points. The data were background corrected using the Bright Source Model for the RXTE PCA\footnote{http://rxte.gsfc.nasa.gov/docs/xte/whatsnew/calibration.html}. The soft colour (SC) was defined as the count rate ratio between PCA Standard 2 spectral channels 7-14 and 3-6 (3.6-6.5 and 2.2-3.6 keV respectively) and the hard colour (HC) as the ratio between channels 21-40 and 15-20 (8.7-16.0 and 6.5-8.7 keV). The intensity was taken to be the 2-60 keV (channel 1-129) count rate.

We checked the colours for instrumental drifts caused by changes in the response of the RXTE PCA by analysing 4 Crab observations covering the entire observation period of the data we used. We created CDs of the Crab observations using the same bands as we used for the GX 13+1 data and calculated the average colours. The shifts of Crab in the CD are below 1\% in SC and below 0.5\% in HC. Dead time effects result in shifts of the data points in the CD relative to one another of less than $\sim$0.1\% in both HC and SC. Both the effect of instrumental drifts and of dead time are negligible compared to the effects reported below.

\section{Analysis and results}
\label{analysisandresults}
\subsection{The colour-colour and hardness-intensity diagrams}
\label{obsandanalysis_cd_hids}
\begin{figure*}[htbp]
\begin{minipage}[t]{0.6\linewidth}
\begin{flushleft}
\includegraphics[height=10.5cm]{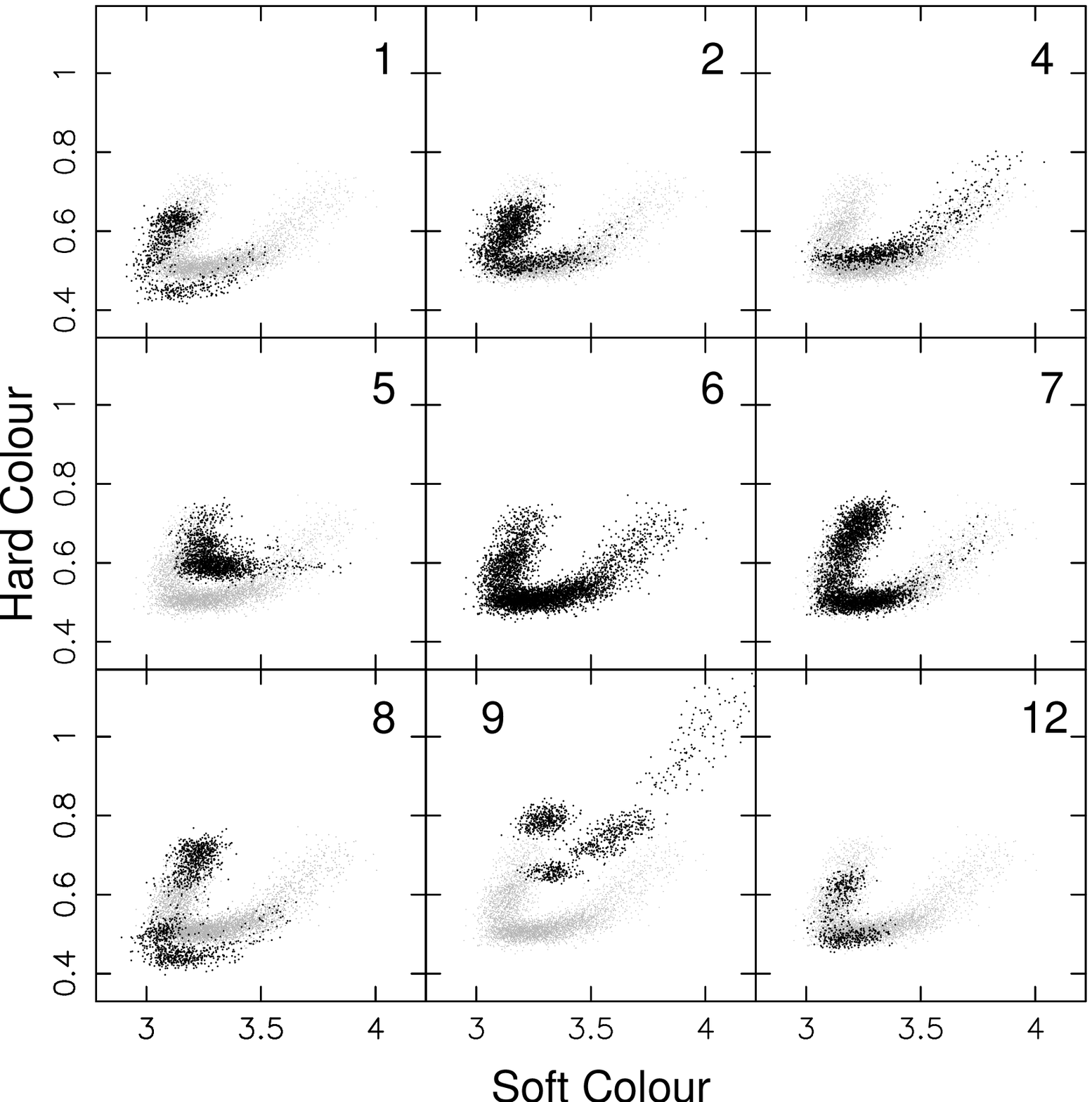}
\end{flushleft}
\end{minipage}%
\begin{minipage}[t]{0.4\linewidth}
\begin{flushleft}
\includegraphics[height=10.5cm]{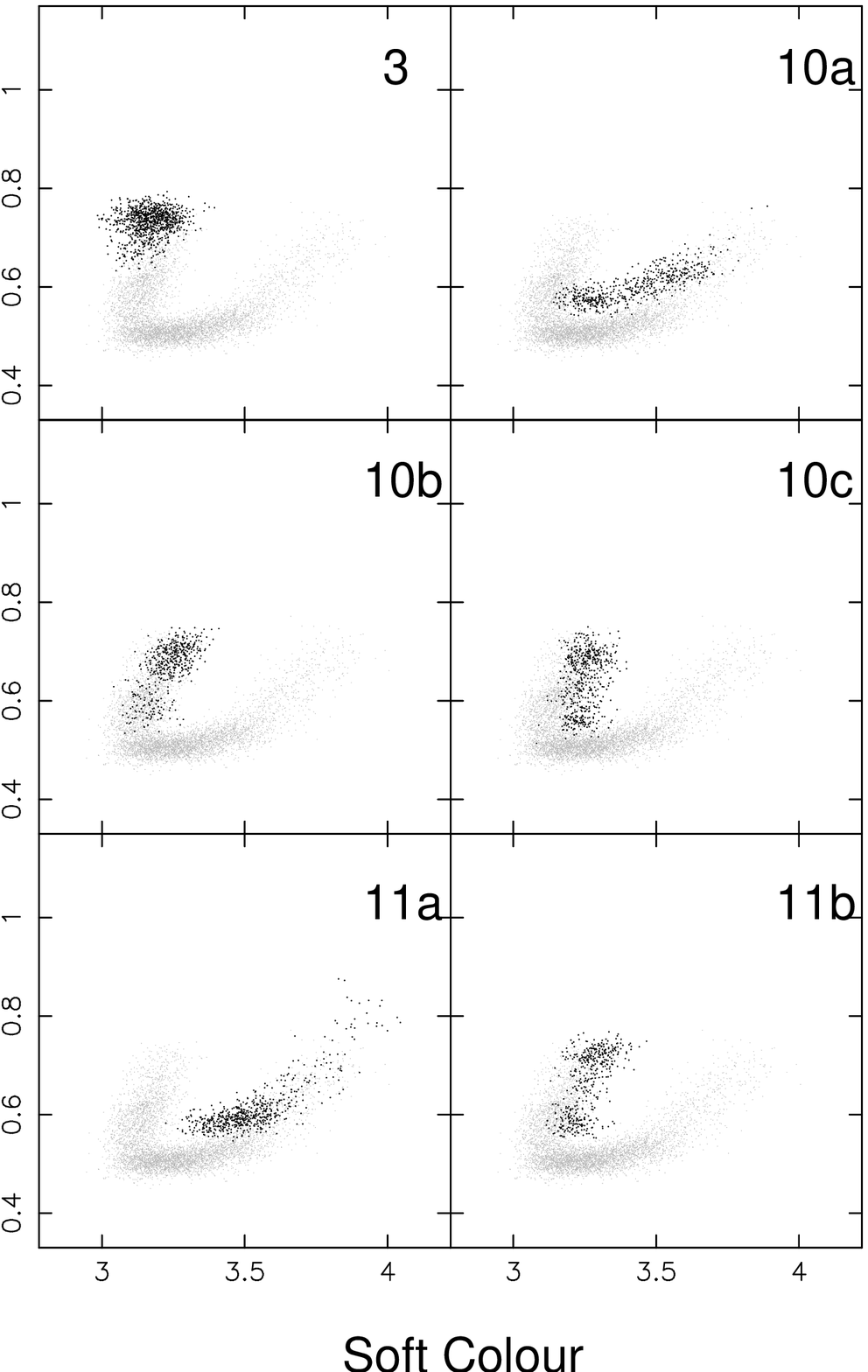}
\end{flushleft}
\end{minipage}
\caption[]{The colour-colour diagrams of tracks 1 to 12 (black points); the tracks that show a vertex are plotted in the left plot and those that do not in the right. For comparison the colour-colour diagram of track 6 is plotted (grey points) in every frame. Notice the shifts of the track, mainly in the hard colour. For each point we used a 16 s average spectrum to calculate the SC (3.6-6.5 keV/2.2-3.6 keV) and HC (8.7-16.0 keV/6.5-8.7 keV). Typical errors are 0.04 (top left of the CD) up to 0.06 (top right of the CD) in the SC and around 0.013 in the HC.}
\label{shifted_tracks}
\end{figure*}

On cursory examination, the CD of our data appears to exhibit the lower part of an atoll track or perhaps the NB/FB part of a Z track, traced out on a time scale of hours to days (Fig. \ref{shifted_tracks}) with a rather sharp vertex. This track shifts on longer time scales, resulting in a CD with overlapping tracks. Similar shifts are also observed in the HIDs. We divided the data into 15 different tracks by adding observations together in chronological order for at most one week, unless the track shifted before then (see Table \ref{ptracks}). A shift in this context was defined as motion of the track over a colour-colour interval more or less perpendicular to the track, significant compared to the track width. Clearly, a shift parallel to the tracks can be overlooked in this way. For that reason data from our 15 tracks could still be to some extent mixed up. However, it is not likely that this happened very often. The reason for this is, that, as shown below, the shifts tend to occur along one particular diagonal in the CD. The tracks do not usually run parallel to that diagonal.

Tracks 3, 10a,b,c and 11a,b were excluded from further analysis, because they do not show a vertex. Because of this lack of a vertex we cannot with confidence compare the positions of these tracks in the CD with those of the others.

\begin{figure}[htbp]
\includegraphics[width=8cm]{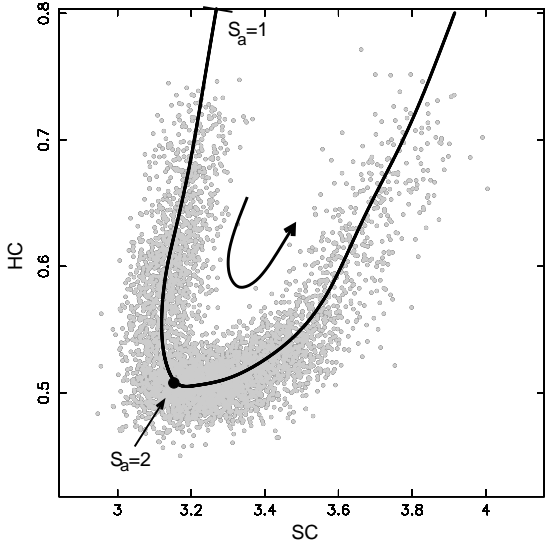}
\caption[]{The CD of track 6 and the spline used to calculate $S_\mathrm{a}$. The locations on the curve where $S_\mathrm{a}=1$ and $S_\mathrm{a}=2$ are indicated; the arrow indicates the direction of increasing $S_\mathrm{a}$.}
\label{sa2point}
\end{figure}

For each of the remaining 9 tracks we defined a parameter $S_\mathrm{a}$ that indicates the position along the track, in a similar way as e.g. \citet{dieters:2000} for the Z source Sco X-1. This was done by approximating the track with a spline onto which the data points are projected along a vector defined by their errors in HC and SC (see Fig. \ref{sa2point}). Different from e.g Sco X-1, GX 13+1 only shows one vertex in the CDs instead of two. We chose the value of $S_\mathrm{a}$ at this vertex to be equal to 2, and fixed the arc length along the track by setting $S_\mathrm{a}$ =1 at a point on the top branch at a fixed HC and SC interval ($\Delta$HC=0.3, $\Delta$SC=0.2) above and to the right of the vertex. The locations of the vertices (the middle of the sharp turn) were estimated by eye. To quantify the shift of each track in the CD we defined a parameter $S_\mathrm{shift}$ as the distance of the track's vertex to the point HC=0.46 and SC=3.07 (the vertex of track 1): $S_\mathrm{shift}\equiv\sqrt{(\Delta HC^2+\Delta SC^2)}$.

\begin{figure}[htbp]
\resizebox{\hsize}{!}{\includegraphics[width=8cm]{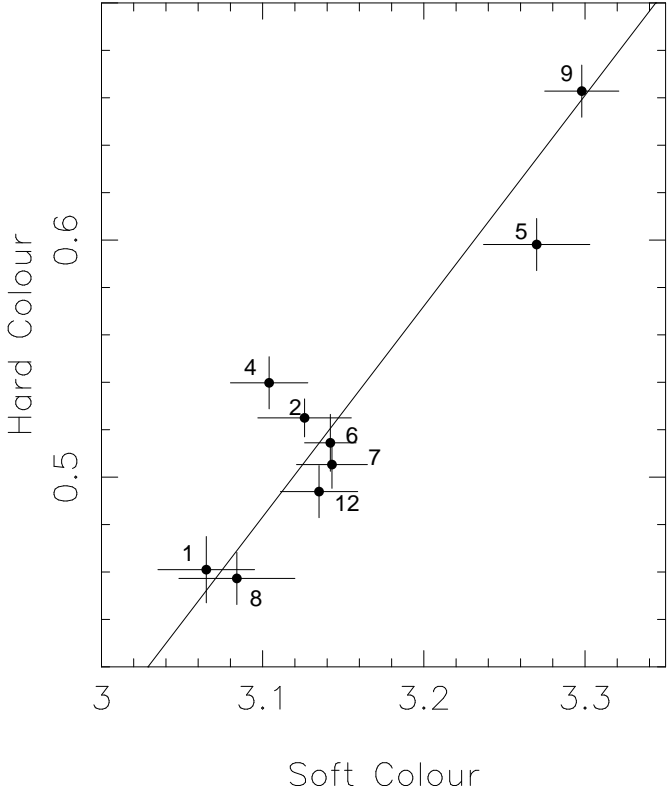}}
\caption[]{The vertex locations of the tracks with their track number and the best linear fit to the points. The vertices are consistent with being on a line of increasing hard and soft colour ($\chi^2/dof = 8.3/7$), spanning a 7\% increase in SC and a 43\% increase in HC.}
\label{vertices1}
\end{figure}

In Fig. \ref{shifted_tracks} we show the CDs of all 15 tracks in black superimposed on track 6 (in grey) for reference. As mentioned above GX 13+1 shows a pattern very reminiscent of a standard atoll track, with, however, a rather sharp vertex. The tracks are shifted relative to each other; the vertex locations are presented in Fig. \ref{vertices1} and Table \ref{ptracks}. The shifts are consistent ($\chi^2/dof = 8.3/7$) with being on a diagonal line in the CD.

\FloatBarrier

\begin{figure*}[htbp]
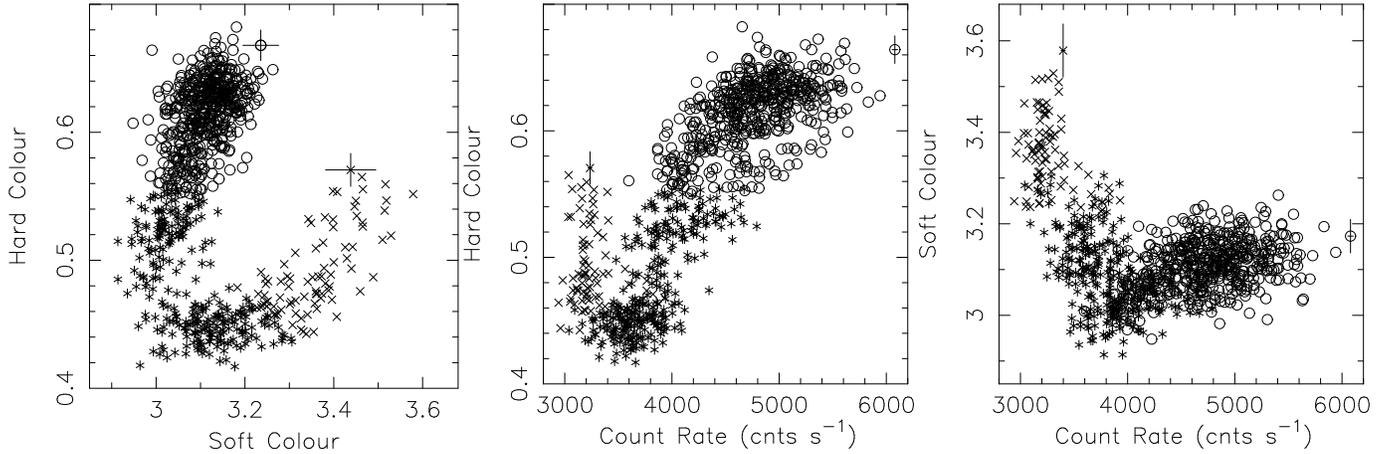

\includegraphics[width=0.33\hsize, angle=-90]{H4165F5a.ps}
\includegraphics[width=0.33\hsize, angle=-90]{H4165F5b.ps}
\includegraphics[width=0.33\hsize, angle=-90]{H4165F5c.ps}
\caption[]{The colour-colour diagram and hardness-intensity diagrams of track 1.
The colour-colour points have different symbols based on their location in the CD to simplify tracking the differences in the shape of the track. The count rates are for 5 PCUs. Typical error bars are shown for two points at the ends of the track in the different diagrams.}
\label{track1_cd&hid}
\end{figure*}

Representative HIDs of track 1 are shown in Fig. \ref{track1_cd&hid}, together with the CD. The colour-colour points have different symbols based on their location in the CD to simplify tracking the differences in the shape of the track between the diagrams. Contrary to what is seen in other sources the track is flipped in the HIDs compared to the CD.

\subsection{Dependence of the count rate on $S_\mathrm{a}$ and $S_\mathrm{shift}$}
\label{flux_sa}
\begin{figure*}[htbp]
\begin{minipage}[t]{0.5\linewidth}
\begin{flushleft}
\includegraphics[width=\linewidth]{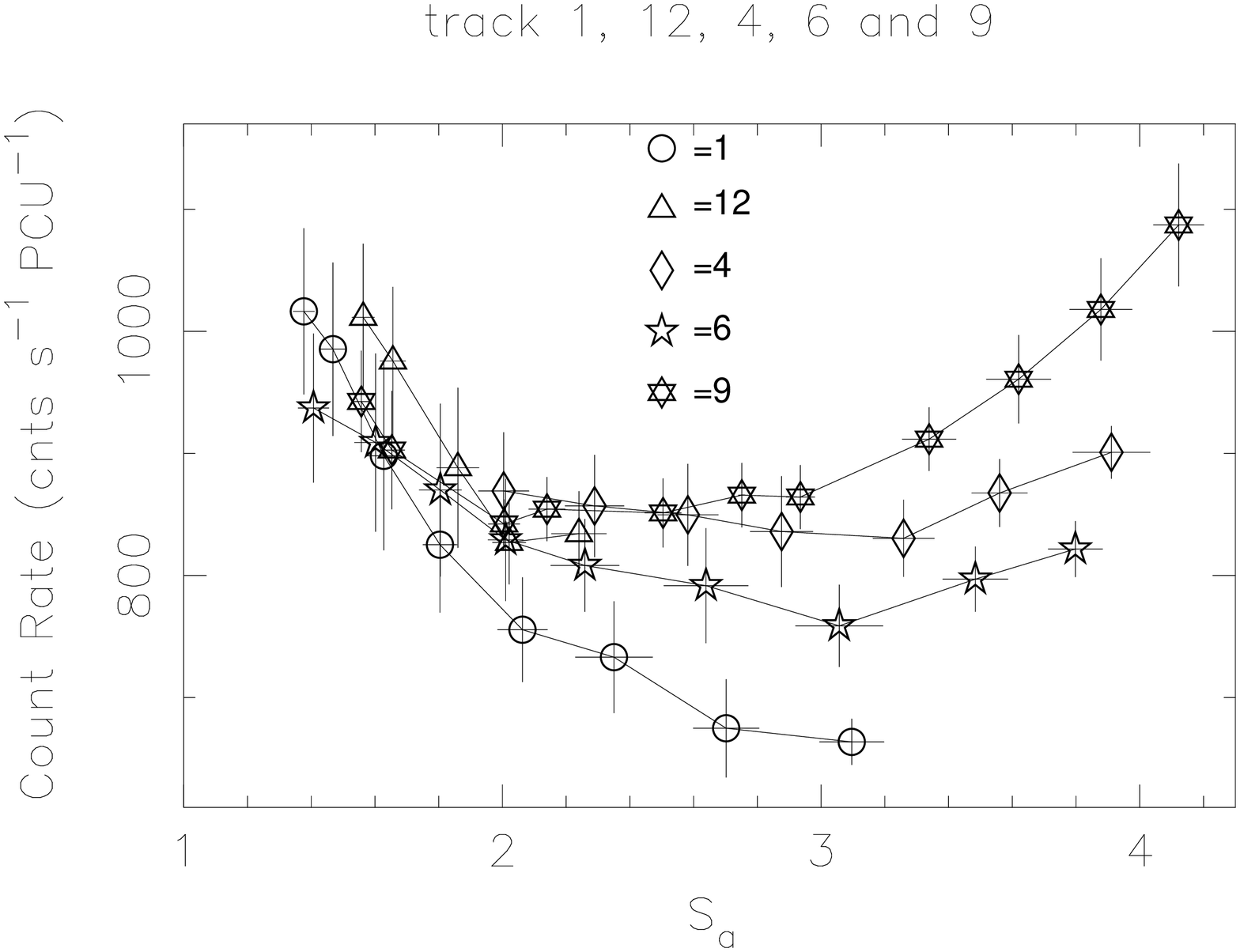}
\end{flushleft}
\end{minipage}%
\begin{minipage}[t]{0.5\linewidth}
\begin{flushright}
\includegraphics[width=\linewidth]{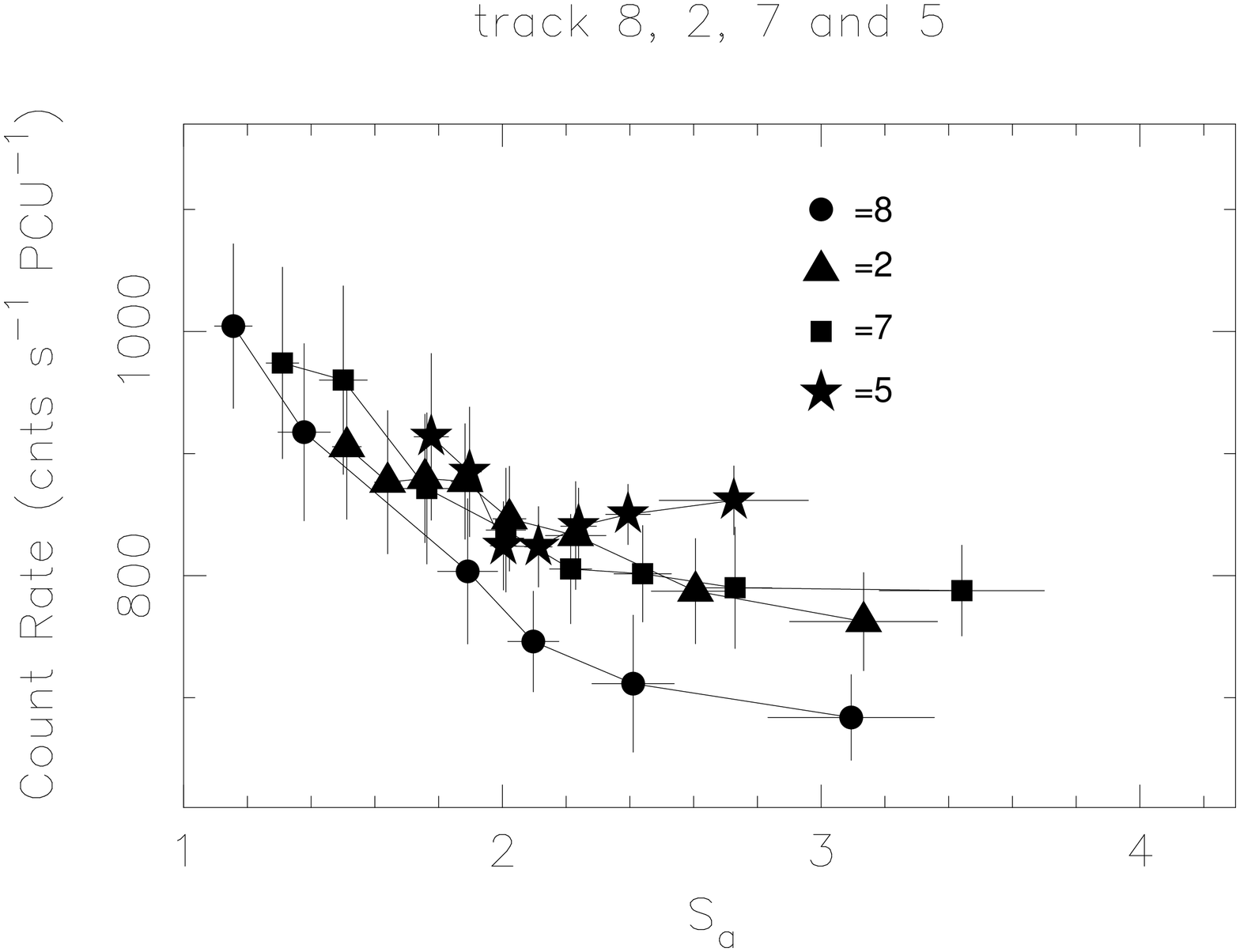}
\end{flushright}
\end{minipage}
\caption[]{The count rate vs $S_\mathrm{a}$ for all the 9 tracks, distributed over two plots for clarity. The track number and corresponding symbol are shown in the figure. For all tracks the count rate decreases up to $S_\mathrm{a}$=2--3.4 and in some tracks it increases again for $S_\mathrm{a}>$2--3.4.}
\label{countrate}
\end{figure*}

\begin{figure}[htbp]
\includegraphics[height=8cm, angle=-90]{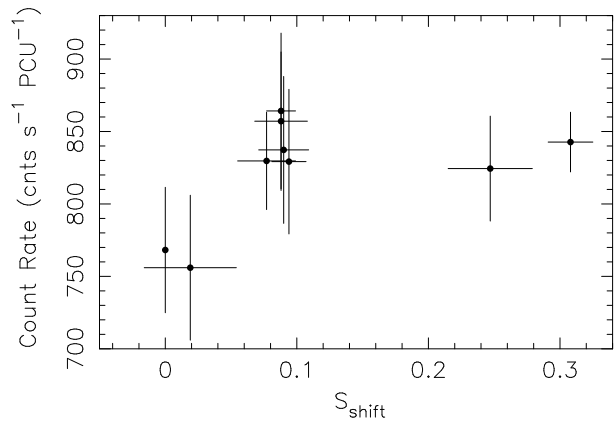}
\caption[]{The count rate at $S_\mathrm{a}=2$ for all 9 tracks plotted vs. $S_\mathrm{shift}$. The two tracks with the lowest $S_\mathrm{shift}$ (tracks 1 and 8) show the lowest count rates.}
\label{fluxsa2}
\end{figure}

In Fig. \ref{countrate} we show the 2-60 keV count rate plotted versus $S_\mathrm{a}$ for each track. Contrary to what is normally seen in atoll sources on these timescales \citep{michiel:1989,tiziana:2001,tiziana:2002}, the count rate depends non-monotonically on $S_\mathrm{a}$, with usually a minimum somewhere in the range $S_\mathrm{a}$=2--3.4. Such a minimum is seen at the NB-FB vertex ($S_\mathrm{a}$=2) in some Z sources \citep[see, e.g, ][]{dieters:2000,jonker:2000,jeroen17+2:2002}, but there the minimum always occurs at the vertex, while in GX 13+1 its position drifts between $S_\mathrm{a}$ values of 2 and 3 depending on observation. In Fig. \ref{fluxsa2} we have plotted the average count rate for $S_\mathrm{a}=2.1\pm0.1$ versus $S_\mathrm{shift}$. From Figs. \ref{countrate} and \ref{fluxsa2} a correlation between count rate and $S_\mathrm{shift}$ is observable: the two tracks with the lowest $S_\mathrm{shift}$, tracks 1 and 8, have lower count rates for a given $S_\mathrm{a}$ than the other tracks, except for the lowest three $S_\mathrm{a}$ intervals of track 1. Tracks 9 and 5 which have the highest $S_\mathrm{shift}$ show the highest count rates above $S_\mathrm{a}\sim2.5$.

\subsection{The power spectra}
\label{thepowerspectra}
\begin{figure*}[htbp]
\begin{flushleft}
\includegraphics[width=0.33\hsize]{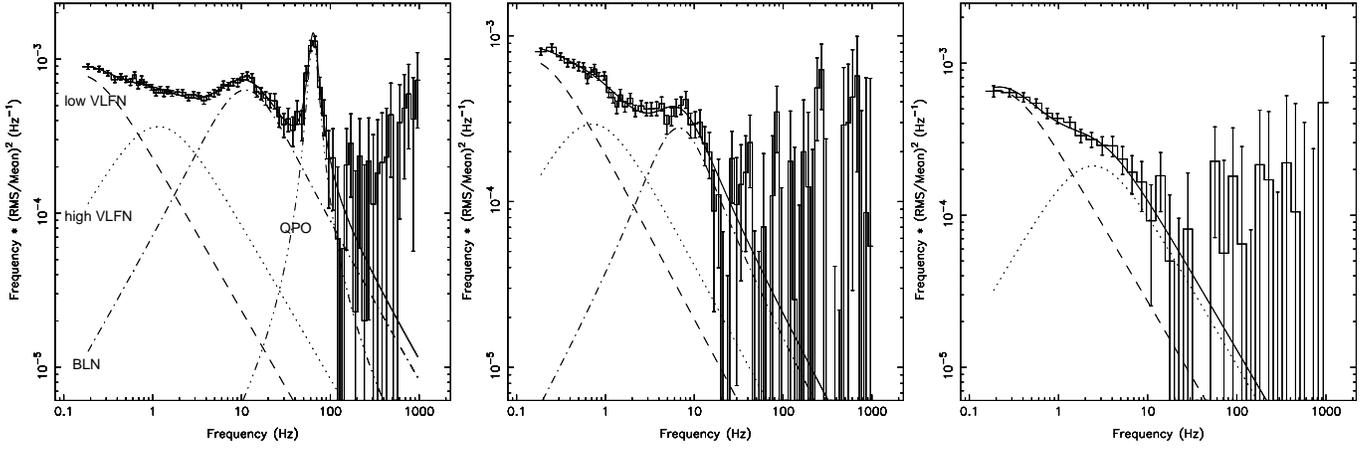}
\includegraphics[width=0.33\hsize]{H4165F8b.ps}
\includegraphics[width=0.33\hsize]{H4165F8c.ps}
\end{flushleft}
\caption[]{Representative power spectra of GX 13+1. Shown are power spectra of track 7 with an averaged $S_\mathrm{a}$ of $1.50\pm{0.08}$ (left; $\chi^2/dof=125/145$), $2.01\pm{0.06}$ (middle; $\chi^2/dof=135/148$) and $2.73{\pm0.12}$ (right; $\chi^2/dof=138/151$). All power spectral features become weaker with increasing $S_\mathrm{a}$. The solid line represents the fit and the dashed and dotted lines the Lorentzian components. The VLFN is fitted by the long dashed line (low VLFN) and the dotted line (high VLFN), the BLN by the dash-single dot line and the QPO by the dash-triple dot line. In the left power spectrum some excess power is present in the 200-2000 Hz range, but no significant kHz QPO is detected.}
\label{example_power_spectra}
\end{figure*}

Power spectra with a Nyquist frequency of 2048 Hz were calculated using 16 s segments of high time resolution data and all energy channels, i.e., covering the 2-60 keV PCA band. Each of the 9 tracks separately was divided into (typically 8) $S_\mathrm{a}$ intervals and the power spectra corresponding to the selected colour-colour points were averaged. A constant Poisson level determined by using the 1024 to 2048 Hz range (where no kHz QPOs were detected) was subtracted. Three representative power spectra from track 7 are presented in power$\times$frequency vs. frequency representation in Fig. \ref{example_power_spectra}. The VLFN, BLN and a QPO are clearly visible.

The power spectra were fitted with 2--4 Lorentzians. We chose to fit the VLFN with two zero-centered Lorentzians instead of a power law, because it has a ``bumpy'' structure, which can not be fitted well with a single power law (to obtain a good fit one has to add additional components to fit the bumps in any case) and is too broad a feature to be well fitted by a single Lorentzian. We will call these two components the low and high VLFN. This approach has the additional advantage that it avoids the common problem that the VLFN power law attempts to also fit power at high frequencies, which is probably unrelated to the VLFN, and can bias the power law to ill represent the data at low frequencies. Two more Lorentzians fit the BLN and a QPO, when present. The Lorentzians are reported in the $\nu_\mathrm{max}$, $Q$, r representation used  by \citet{belloni:2002}, where $\nu_\mathrm{max}$ is the frequency at which the Lorentzian reaches its maximum in a power$\times$frequency vs. frequency power spectral representation, $Q$ the quality factor (for which we use the standard definition of centroid frequency divided by full width at half maximum) and r the fractional rms amplitude. We will regard $\nu_\mathrm{max}$ to be the characteristic frequency of a Lorentzian component. The rms powers were calculated in the 0--$\infty$ Hz range, except for the two components fitting the VLFN where we used the 0.125--$\infty$ Hz range because for these components the unmeasured power below our lowest Fourier frequency provides a significant contribution (which is not the case for the other components). The rms amplitudes were calculated assuming an average background of 140 cnts $\mathrm{s^{-1}}$ (2-60 keV, 5 PCUs); the $\sim$20 cnts $\mathrm{s^{-1}}$ deviations from this average are negligible in view of the high source count rates.

In all of the tracks we have found VLFN and BLN and in 5 of the 9 tracks we detected the 57-69 Hz QPO first reported by \citet{jeroen:1998}. For all $S_\mathrm{a}$ intervals we found VLFN; the BLN usually disappeared in the highest $S_\mathrm{a}$ intervals and QPOs were only detected in the lowest $S_\mathrm{a}$ intervals. The power spectral components and their dependence on $S_\mathrm{a}$ and $S_\mathrm{shift}$ is discussed in more detail in Sect. \ref{thepsp}.

\subsection{Dependence of the power spectral properties on $S_\mathrm{a}$ and $S_\mathrm{shift}$}
\label{thepsp}
\subsubsection{The very low frequency noise}
\begin{figure*}[htbp]
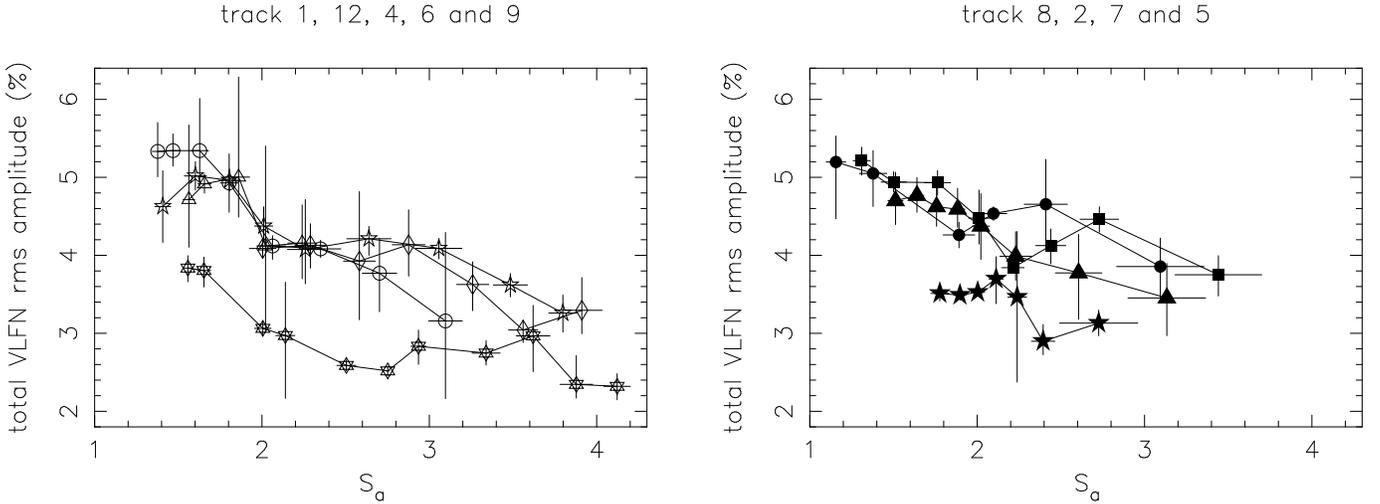

\begin{minipage}[t]{0.5\linewidth}
\begin{flushleft}
\includegraphics[height=0.95\linewidth, angle=-90]{H4165F9a.ps}
\end{flushleft}
\end{minipage}
\begin{minipage}[t]{0.5\linewidth}
\begin{flushright}
\includegraphics[height=0.95\linewidth, angle=-90]{H4165F9b.ps}
\end{flushright}
\end{minipage}
\caption[]{VLFN fractional rms amplitude vs $S_\mathrm{a}$ for all 9 tracks; the amplitude plotted is that of the combined low and high VLFN, integrated from 0.125 to $\infty$ Hz. For all tracks the amplitude of the VLFN generally decreases with $S_\mathrm{a}$, and tracks 5 (five-pointed stars, right frame) and 9 (six-pointed stars, left frame), which have the highest $S_\mathrm{shift}$, have the lowest amplitude for a given $S_\mathrm{a}$. Symbols are the same as in Fig. \ref{countrate}.}
\label{VLFN_power}
\end{figure*}

As described in Sect. \ref{thepowerspectra}, the VLFN was fitted with two zero-centered ($Q=0$) Lorentzians. The power of both these Lorentzians was added to calculate the total rms amplitude of the VLFN. For all 9 tracks the rms of the total VLFN generally decreases with $S_\mathrm{a}$ (see Fig. \ref{VLFN_power}), from $\sim$4.5--5.5\% to $\sim$3--3.5\%. An anti-correlation with $S_\mathrm{shift}$ is visible: tracks 5 and 9, which have a high $S_\mathrm{shift}$, have a VLFN rms that is significantly lower than that of the other tracks. To quantify this we fitted a straight line to the VLFN amplitude - $S_\mathrm{a}$ relation, and determined the value of the VLFN amplitude at $S_\mathrm{a}=2$ from this fit. In Fig. \ref{VLFNsa2} we show this amplitude as a function of $S_\mathrm{shift}$. As $S_\mathrm{shift}$ increases from 0 to $\sim$0.3 the VLFN rms at $S_\mathrm{a}=2$ decreases from $\sim$4.8\% to $\sim$3\%.

The absolute rms of the VLFN as a function of $S_\mathrm{a}$ of all tracks combined is shown in Fig. \ref{absoluteVLFN_power}. Up to $S_\mathrm{a}=2$ the absolute rms decreases from $\sim$50 to 30 cnts $\mathrm{s^{-1}}$, and for $S_\mathrm{a}>2$ it stays approximately constant at $\sim$30 cnts $\mathrm{s^{-1}}$. So for $S_\mathrm{a}<2$ the decreasing absolute rms is correlated to the decreasing count rate; however, for $S_\mathrm{a}>2$ the count rate in some tracks increases, but this does not affect the absolute rms, which stays approximately constant.

The characteristic frequency ($\nu_\mathrm{max}$) of the low and high VLFN does not show a clear correlation with $S_\mathrm{a}$, but varies between 0.1--0.3 Hz for the low and between 0.5--1.5 Hz for the high VLFN (with a few exceptions up to 2 Hz when no BLN was fitted, and for low $S_\mathrm{a}$ in track 9 where the BLN also shows an extremely high $\nu_\mathrm{max}$).

\subsubsection{The band limited noise}
\label{bln}
\begin{figure*}[htbp]
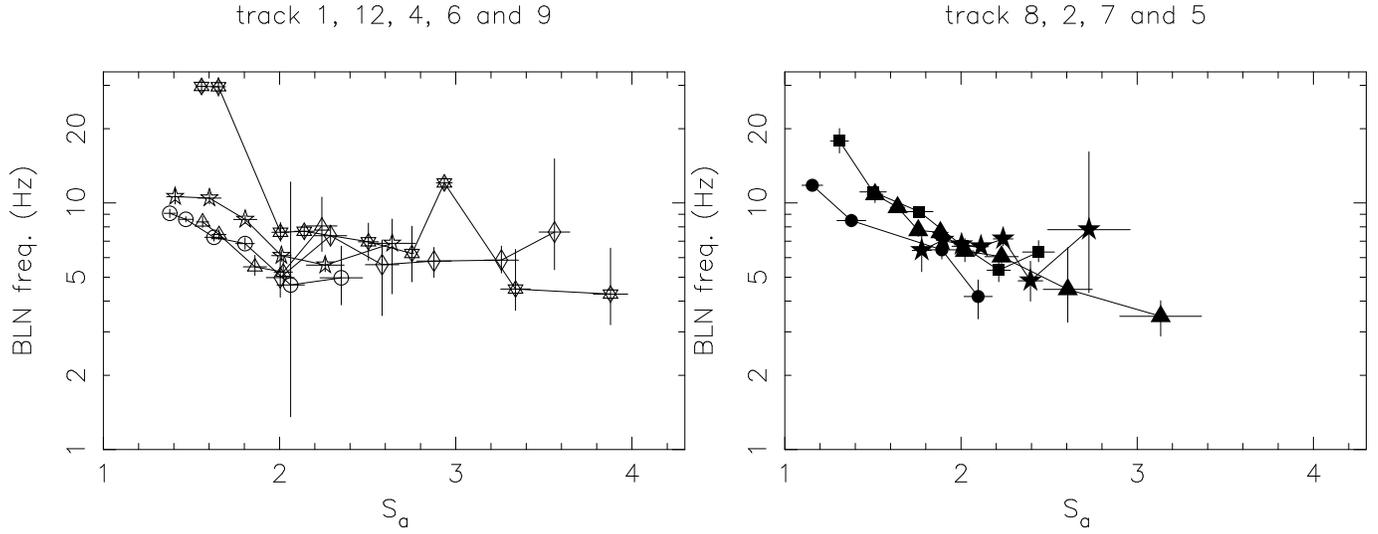

\begin{minipage}[t]{0.5\linewidth}
\begin{flushleft}
\includegraphics[height=\linewidth, angle=-90]{H4165F12a.ps}
\end{flushleft}
\end{minipage}
\begin{minipage}[t]{0.5\linewidth}
\begin{flushright}
\includegraphics[height=\linewidth, angle=-90]{H4165F12b.ps}
\end{flushright}
\end{minipage}
\caption[]{The characteristic frequency of the BLN vs $S_\mathrm{a}$ for all 9 tracks. The characteristic frequency generally decreases with $S_\mathrm{a}$. Track 9 shows an extremely high frequency for its lowest two $S_\mathrm{a}$ intervals. Symbols are the same as in Fig. \ref{countrate}; in some $S_\mathrm{a}$ intervals no BLN was detected.}
\label{BLN_freq}
\end{figure*}

\begin{figure*}[htbp]
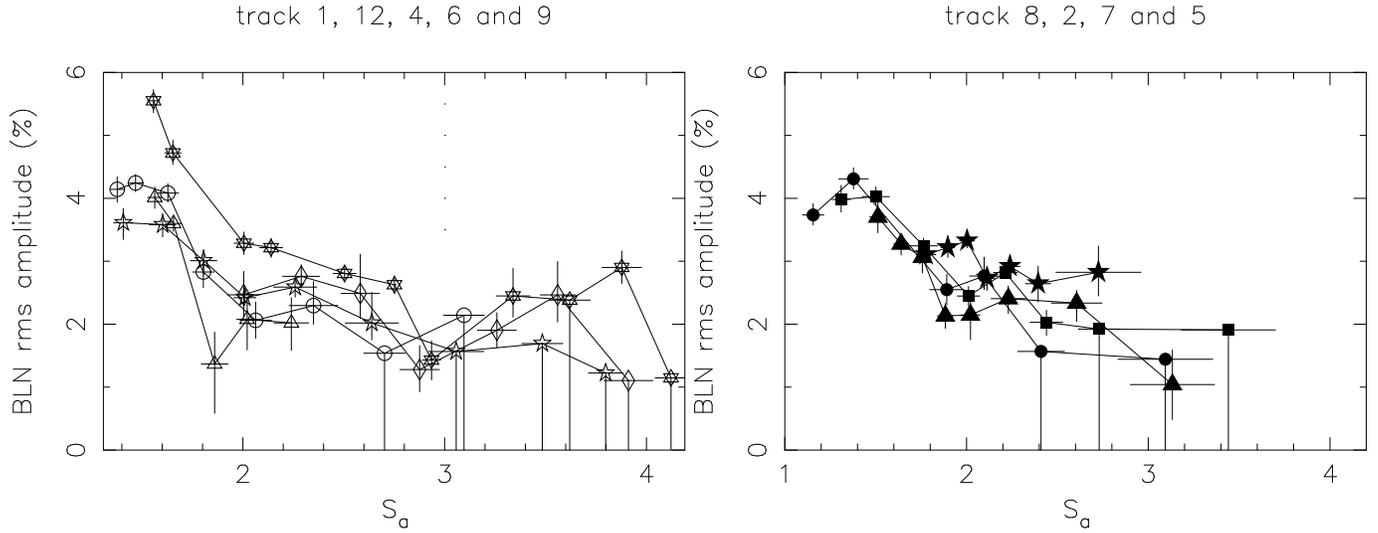

\begin{minipage}[t]{0.5\linewidth}
\begin{flushleft}
\includegraphics[height=\linewidth, angle=-90]{H4165F13a.ps}
\end{flushleft}
\end{minipage}
\begin{minipage}[t]{0.5\linewidth}
\begin{flushright}
\includegraphics[height=\linewidth, angle=-90]{H4165F13b.ps}
\end{flushright}
\end{minipage}
\caption[]{The fractional rms amplitude of the BLN vs $S_\mathrm{a}$ for all 9 tracks. The amplitude of the BLN generally decreases with $S_\mathrm{a}$ for all tracks, but in track 4 (diamonds, left frame) and 9 (six-pointed stars, left frame) it increases again for the highest two $S_\mathrm{a}$ intervals. $3\sigma$ Upper limits were determined with $Q=0.5$ and $\nu_\mathrm{max}=5$ Hz, and are represented by the points with only a a negative error bar extending to the bottom of the plot. Symbols are the same as in Fig. \ref{countrate}}
\label{BLN_pow}
\end{figure*}

The BLN characteristic frequency, shown in Fig. \ref{BLN_freq}, and rms, shown in Fig. \ref{BLN_pow}, both decrease with $S_\mathrm{a}$. For most tracks the frequencies decrease from $\sim$10 to $\sim$4--5 Hz, except for track 9 which has a BLN frequency of $\sim$30 Hz for its lowest two $S_\mathrm{a}$ intervals ($S_\mathrm{a}\sim$1.5--1.7).
The highest BLN rms is $\sim$4--6\% and is found for low $S_\mathrm{a}$; when $S_\mathrm{a}$ increases the BLN rms decreases until it becomes undetectable around $S_\mathrm{a}\sim$2.5--3. Typical 95\% confidence upper limits for the BLN for the $S_\mathrm{a}$ intervals where it was not detected are $\sim$2\%, with $\nu_\mathrm{max}=5$ Hz and $Q=0$.5. For tracks 9 and 4 the rms increases again with $S_\mathrm{a}$ above $S_\mathrm{a}\sim3$. 

The characteristics of the BLN do not show a clear correlation with $S_\mathrm{shift}$, although the BLN in track 9 has a higher rms for a given $S_\mathrm{a}$ than in the other tracks and also shows extremely high frequencies for low $S_\mathrm{a}$ (see above). The $Q$ of the BLN is also not significantly correlated with $S_\mathrm{a}$. It varies between 0 and 1, with the exception of the last interval of track 2 ($S_\mathrm{a}=3.13\pm0.23$, $Q=2.6^{+22.1}_{-1.7}$) and $S_\mathrm{a}$ interval 7 of track 9 ($S_\mathrm{a}=2.94\pm0.05$, $Q=4^{+4}_{-2}$), where the BLN also shows a very high frequency of $12.1^{+0.7}_{-0.8}$ Hz; in these two cases (only) this component is sufficiently narrow to be called a QPO.

\subsubsection{Quasi-periodic oscillations}
\label{qpo_section}
QPOs were detected with frequencies ranging from $\sim$60--75 Hz and $Q$ values of $\sim$2--5 (detailed properties are shown in Table \ref{qpo_prop}). They are only detected at low $S_\mathrm{a}$ values, with the highest rms at the lowest $S_\mathrm{a}$, and in too small a range of $S_\mathrm{a}$ to say anything about the dependence of their frequency and $Q$ on $S_\mathrm{a}$. The fact that QPOs were detected in tracks 1 and 8 and not in tracks 5 and 9 could be related to the high $S_\mathrm{shift}$ values of tracks 5 and 9, but this could also be due to the fact that these tracks do not reach the upper-left most part of the track, where the QPOs are found. The frequencies of the detected QPOs are all consistent with being in the range of those previously found by \citet{jeroen:1998}.

We have searched for kHz QPOs up to 2048 Hz, but none were detected above the 3$\sigma$ confidence level. The strongest narrow features ($Q\sim10$) above 300 Hz were between 2 and 3$\sigma$ and resulted in 3$\sigma$ upper limits of up to 3\% rms. A broad, 3.1$\sigma$, high frequency feature ($2.6\pm0.5$ \% rms, $Q=0.9^{+0.9}_{-0.6}$, $\nu_\mathrm{max}=858^{+225}_{-137}$ Hz) is found in track 7 ($S_\mathrm{a}=1.50\pm0.08$), but when the number of trials is taken into account this is not a significant detection.

\FloatBarrier

\FloatBarrier

\FloatBarrier

\FloatBarrier

\begin{figure*}[htbp]
\begin{minipage}[t]{0.5\linewidth}
\begin{flushleft}
{\includegraphics[height=0.95\linewidth, angle=-90]{H4165F10.ps}}
\end{flushleft}
\caption[]{The VLFN amplitude at $S_\mathrm{a}$=2 for all 9 tracks (as determined from a linear fit to a plot of the VLFN rms vs. $S_\mathrm{a}$), as a function of $S_\mathrm{shift}$. The amplitude decreases from $\sim$4.8\% to $\sim$3\% as $S_\mathrm{shift}$ increases from 0 to $\sim$0.3.}
\label{VLFNsa2}
\end{minipage}
\begin{minipage}[t]{0.5\linewidth}
\begin{flushright}
{\includegraphics[height=0.95\linewidth, angle=-90]{H4165F11.ps}}
\end{flushright}
\caption[]{The absolute amplitude of the VLFN as a function of $S_\mathrm{a}$ for all 9 tracks combined. Up to $S_\mathrm{a}$=2 the absolute rms amplitude decreases from $\sim$50 to 30 cnts $\mathrm{s^{-1}}$, and for $S_\mathrm{a}>$2 it stays approximately constant at $\sim$30 cnts $\mathrm{s^{-1}}$.}
\label{absoluteVLFN_power}
\end{minipage}
\end{figure*}

\FloatBarrier

\clearpage
\section{Discussion}
\subsection{Extreme secular motion in CD and HIDs}
We have studied the secular motion both in the CD and in the HIDs of GX 13+1. The largest shifts are observed in the direction of HC ($\sim$40\%) and not in SC ($\sim$7\%). In other atoll and Z sources which show secular motion, shifts of less than 3\% in HC and $\sim$4\% in SC \citep[4U 1636-53, ][]{tiziana:2002} and 6\% in HC and 8\% in SC \citep[GX 5-1, ][]{kuulkers:1994} are observed; in Cyg X--2 the shifts in both HC and SC are up to $\sim$10\% \citep{Kuulkers:1996}. This is not strongly dependent on the choice of bands. The shifts in the CD of GX 13+1 are remarkably large, particularly in HC, even when compared to Z sources, where the shifts are larger than in atoll sources. As is also seen in other sources (references above), the shift of the track shows a correlation with average count rate resulting in shifted tracks in the HIDs. We find that in GX 13+1 the amplitude of the VLFN also shows a correlation with this shift; it decreases when the track is shifted towards higher HC and SC. Such a correlation is not observed in the Z source GX 5-1, but in the atoll source 4U 1636-53 the part of the track that shifted to higher SC and intensity also has a lower VLFN rms. The VLFN rms in Cyg X--2 shows the same dependence on the shift of the track, but the overall intensity level of a track decreases with this shift \citep{kuulkers:1999}. In GX 13+1, track 9, which has the largest shift also has unusual BLN properties (Sect. \ref{bln}); in 4U 1636-53 an extra, broad component ($Q=0.8$) is detected next to the BLN in the high intensity part of the track at $\sim1.1\%$ rms, but not in the low intensity part (with $1\sigma$ upper limits of $\sim$0.6\%).

\subsection{The flipping effect}
In the HIDs with the HC or SC on the vertical, and count rate on the horizontal axis the arc-shape found in the CD is reversed, i.e., the source moves through the arc in the opposite sense in the HIDs as compared to the CD. This is different from what is usually observed in atoll sources \citep{rudi:1999,steve:2000,muno:2002,tiziana:2002}, where, on short timescales, the same basic pattern is traced out in both the CD and the HIDs and the sense of motion is always the same. On longer timescales secular shifts in the HIDs are sometimes observed \citep{steve:epsilon}. In most Z sources the observed basic pattern in the CD and HIDs is also similar \citep{schulz:1989,Kuulkers340+0:1996,jonker:2002,jeroen17+2:2002}, although for some sources the count rate decreases along (a part of) the FB, causing this branch to be in the opposite direction in the HIDs compared to the CD \citep{Kuulkers:1996}. The cause for this unusual behaviour of GX 13+1 in the HIDs is that the count rate depends on $S_\mathrm{a}$ in a different way than in most other sources (see Sect. \ref{flux_sa}). A possible explanation for this will be discussed in Sect. \ref{explanation}.

\subsection{GX 13+1 in the Z/atoll scheme}
The tracks that GX 13+1 traces out in the CD are, at first sight, very reminiscent of a classic island/banana in an atoll source, or a normal-branch/flaring-branch in a Z source. However, from the results presented in Sect. \ref{analysisandresults} we find that the power spectral properties of this source as a function of $S_\mathrm{a}$ resemble neither those of a classic atoll, nor of a Z source.

In a given track the count rate of GX 13+1 decreases along the track up to $S_\mathrm{a}$=2--3 after which (in some tracks) it increases again with $S_\mathrm{a}$, while in atoll sources on short timescales the count rate generally increases monotonically with $S_\mathrm{a}$ along the whole track \citep[see Sect. \ref{flux_sa}; for references on atoll source behaviour see][]{steve:2002,tiziana:2002,michiel:1989}. This is also different from what is observed in Z sources, where the count rate can slightly decrease along the NB, but increases immediately after the NB/FB vertex so that the minimum in count rate is at the vertex \citep[see Sect. \ref{flux_sa}; for references on Z source behaviour see ][]{dieters:2000,jonker:2000,oneill:2001,jonker:2002,jeroen17+2:2002}. The VLFN of GX 13+1 is very strong (usually $\sim$4--5.5\% rms for $S_\mathrm{a}\leq$2) compared to atoll sources in the island state or lower banana branch and Z sources in the NB (the VLFN frequency range is similar). The VLFN rms of GX 13+1 generally decreases with $S_\mathrm{a}$. This is the opposite of what is normally seen in both atoll and Z sources, where the VLFN increases with $S_\mathrm{a}$ along the whole track \citep[see, however, ][ on the atoll source 4U 1820-30]{rudi:1999}. The BLN is quite weak ($\sim$2--5\%) compared to that in atoll sources in the island state and lower banana and its rms decreases with $S_\mathrm{a}$ as is usual in atoll and Z sources, but in tracks 4 and 9 it increases again in the highest $S_\mathrm{a}$ intervals. The characteristic frequency of the BLN also decreases with $S_\mathrm{a}$, which, in either atoll sources in the island state and lower banana branch or Z sources, has so far only been observed in the atoll source 4U 1820-30, where a BLN-like component that evolves into or is replaced by a $\sim$7 Hz QPO \citep{rudi:1999} is seen. In the Z source GX 349+2, a similar noise component is seen of which the centroid frequency decreases with $S_\mathrm{a}$, but the $\nu_\mathrm{max}$ of this component shows a more complicated dependence on $S_\mathrm{a}$ \citep{oneill:2002}. Generally, in atoll sources the $\nu_\mathrm{max}$ of the BLN increases with $S_\mathrm{a}$ up to the upper banana, where it starts to decrease again, whereas in Z sources it increases monotonically.

We find the QPO properties of GX 13+1 to be unlike either Z or atoll sources. At the lowest $S_\mathrm{a}$ intervals the 57-69 Hz QPO found by \citet{jeroen:1998} is detected. In atoll sources a low frequency QPO is usually detected in the island state with a frequency of 10-15 Hz \citep[see, however, ][]{rudi:1999} ,increasing up to 40-47 Hz towards the upper banana, where it becomes undetectable \citep[although a second low frequency QPO is sometimes also seen at similar frequencies, but for higher $S_\mathrm{a}$, see][]{steve:2002}. If the QPO would remain detectable, one would expect to find such a 57-69 Hz QPO at higher $S_\mathrm{a}$, in the middle banana branch. The HBO in Z sources has a frequency up to $\sim$55 Hz in the upper NB; if the QPO in GX13+1 is interpreted as this QPO one would also expect to find a $\sim$6 Hz NBO on this branch; in the Z source GX 349+2 no NBO is found, but a very broad noise component is seen instead \citep{kuulkers:1998,oneill:2001}, and of this GX 13+1 shows no evidence either. The fact that no kHz QPOs are detected anywhere along the branch (see Sect. \ref{qpo_section}) is uncharacteristic for both atoll and Z sources.

Clearly, GX 13+1 is a peculiar source in which the properties correlate to $S_\mathrm{a}$ in a very unusual way when compared both to atoll and Z sources. In the final section we explore a possible explanation for this unusual behaviour.

\subsection{A jet scenario}
\label{explanation}
If the behaviour of GX 13+1 is interpreted within the framework of atoll source behaviour, then clearly for high $S_\mathrm{a}$ the source is in the upper banana. Based on the CD, at low $S_\mathrm{a}$ one would expect the source  to be in the lower left banana and island state. However, the fact that for low $S_\mathrm{a}$ 57-69 Hz QPOs, and weak BLN with a characteristic frequency that decreases with $S_\mathrm{a}$ are found, suggests that, contrary to what one would expect based on the CD, the ``true'' state of the source there is the middle part of the banana branch, because that is where other atoll sources can be expected to show this kind of power spectral behaviour. This is supported by the fact that so far no kHz QPOs have been found in this source, consistent with their absence in other atoll sources in the middle and upper parts of the banana branch.

In GX 13+1 the count rate decreases as $S_\mathrm{a}$ increases up to $S_\mathrm{a}$=2--3, and only for higher $S_\mathrm{a}$ increases, which is what in a regular atoll source one would expect along the whole track (see Sect. \ref{flux_sa}). One interpretation for this unusual behaviour would be that there is something adding extra flux at low $S_\mathrm{a}$, which becomes weaker towards higher $S_\mathrm{a}$ resulting in a count rate that first decreases, and then increases as the standard atoll upper banana behaviour takes over. From the two branched structure in the CDs one can conclude that this extra flux would have to be hard, because for low $S_\mathrm{a}$ the source spectrum gets harder when the count rate increases. This scenario would also explain why the BLN is so weak and shows a decreasing $\nu_\mathrm{max}$ along the track, and why no kHz QPOs are detected at low $S_\mathrm{a}$: these properties are normal for the middle and upper parts of the banana branch, which is, by hypothesis, the true state of the source. In most atoll sources the low frequency QPO disappears at the same $S_\mathrm{a}$, or a slightly lower value, as the kHz QPO \citep{steve:2002,tiziana:2002}; the fact that in GX 13+1 we still see the low frequency QPO while the kHz QPO is absent is unusual, but in line with the relatively high frequency of the low frequency QPO. So the hypothesis we are exploring is that the hard extra flux creates an anomalous second branch in the CD reminiscent of a lower banana and island branch, while the source remains in the ``middle'' banana state.

A possible source for this extra, hard flux could be a relativistic jet. In general in atoll and Z sources radio emission, ascribed to the presence of a relativistic jet \citep{fender:2001}, tends to increase towards lower $S_\mathrm{a}$. If this more intense jet is associated with the extra hard flux, this could cause a second branch in the CD, while the source is still in the upper banana state. We further hypothesize that the unusually strong VLFN would occur in the hard jet flux. Consequently, when the jet becomes weaker towards higher $S_\mathrm{a}$ this causes the VLFN to decrease in amplitude. This is in accordance with the fact that up to $S_\mathrm{a}$=2 the absolute rms amplitude of the VLFN decreases, from which it can be concluded that it is the flux, that also decreases in this part of the track, that shows relatively strong VLFN that disappears. For $S_\mathrm{a}>$2 the absolute rms amplitude of the VLFN stays approximately constant, while in some tracks the flux increases for $S_\mathrm{a}>$3. The extra flux in these tracks does lead to a significant increase of the absolute rms amplitude.

The question of course is: why would a jet in GX 13+1 cause this peculiar behaviour and not in other atoll sources? The radio emission of GX 13+1 is unusually strong for an atoll source \citep{fender:2000}, so perhaps for a reason unknown the jet is unusually strong in this source. A more attractive possibility is that it is not the jet that is unusual, but the system's orientation. If the jet is pointing directly towards us Doppler boosting could significantly increase its radio flux \citep[e.g.][]{fender:2000}; a strongly relativistic jet \citep[which has already been directly imaged for the Z source Sco X-1, ][]{fomalont:2001a,fomalont:2001b} could easily Doppler boost the radio emission by more than an order of magnitude, putting the radio emission of GX 13+1 in accordance with measurements and upper limits of other atoll sources \citep{fender:2000}. The unusual X-ray phenomena could then for example be the consequence of a better view of the X-ray emitting regions associated with the base of the jet, offered by the unusual orientation of the disk, possibly assisted by Doppler boosting as well. A similar possibility of Doppler-boosted jet X-ray emission causing hard flaring (``microblazar'') was recently explored for Cyg X-1 by \citet{blazar:2002}.

The large shifts of the track in the CD of GX 13+1 compared to other atoll and Z sources might be related to this better view of the X-ray emitting base of the jet region as well. This would require this region to contribute different baseline levels of hard flux at different epochs in order to produce the secular motion (which, consistent with this, takes place mostly in the HC), VLFN whose amplitude anticorrelates, as well as extra flux at $S_\mathrm{a}>3$ which correlates, with this baseline flux level. This could for example be related to secular geometry changes (e.g. precession) modulating our view of the base of the jet, or to variations in the jet activity itself.

One can also attempt to explain the peculiar behaviour of GX 13+1 with a similar relativistic jet scenario assuming the source is a Z source. The QPO observed in GX 13+1 would then be interpreted as the HBO, occurring at the top of the NB, leading to the conclusion that the CD shows both a NB and a FB. The strong VLFN that decreases along the whole track might be related to a jet similarly to the atoll scenario. This jet could be related to the high radio over X-ray luminosity ratio compared to other Z sources. However, this model leaves unexplained the decreasing frequency of the BLN with $S_\mathrm{a}$ and the absence of an NBO.

So, when a choice is made between an atoll or a Z classification for GX 13+1 (in either case modified by an unusual jet geometry) the overall phenomenology favours the former option. An alternative would be that GX 13+1 combines some properties of atoll and Z sources (as noted in Sect. \ref{intro}, GX 13+1 has been described as the atoll source closest to the Z sources). This possibility will be further explored in a forthcoming paper \citep{jeroen:2002}.

It has been suggested that narrow absorption features, as found in GX 13+1 \citep{ueda:2001}, are related to a high inclination of the source, contrary to the low inclination predicted by the scenario discussed above. However, it is also possible that the absorption features are not related to the inclination of the source, but, as previously suggested by \citet{ueda:1998}, to a highly ionized plasma associated with their relativistic jets, causing absorption from the flux from the hot inner accretion region.

Several predictions can be made based on the atoll source+jet hypothesis. Since the unusually strong VLFN occurs in the jet flux, one expects the VLFN, at low $S_\mathrm{a}$, to be the strongest in the hard flux. Also, as the jet is the main source of radio emission in LMXBs, we expect the radio emission to be strongest in the upper part of the second branch in the CD (at low $S_\mathrm{a}$), decreasing as the source moves towards the vertex.  This can be tested with simultaneous radio and X-ray observations. One would have to account for a delay factor in the radio emission of perhaps up to half an hour \citep{mark:2002} associated with the expansion of the synchrotron clouds. 
Such observations have been done by \citet{garcia:1988}, but with a small time range ($\sim$5 hours of overlapping X-ray and radio observations) and no CD analysis was performed, and by \citet{berendsen:2000}, also with a short period of overlapping X-ray and radio observations. Recent simultaneous radio and X-ray observations also exist \citep[to be reported in ][]{jeroen:2002}, and preliminary results of this work that became available after the current analysis and interpretation were completed, appear to confirm our suggestion. 

{\acknowledgements RS and MK acknowledge support from the Netherlands Organisation for Scientific Research (NWO).}
\bibliographystyle{aa}
\bibliography{H4165prepr}

\newpage

\begin{table*}
\begin{flushleft}
\begin{tabular}{cccccccccc}
\hline
\hline
Track&\multicolumn{3}{c}{Vertex}&Flux at $S_\mathrm{a}=2$&\multicolumn{2}{c}{Observation ID}&$t_\mathrm{start}$ (UTC)&$t_\mathrm{end}$ (UTC)&$t_\mathrm{obs}$\\
nr&HC&SC&$S_\mathrm{shift}$&(cnts s$^{-1}$ PCU$^{-1}$)&\multicolumn{2}{c}{of included observations}&\multicolumn{2}{c}{month-day hours:min}&(ks)\\
\hline
1&0.46&3.07&0.00&768.2$\pm$43.3&30051-01-01-01&30051-01-01-00&05-17 11:01&05-17 20:57&28\\
2&0.53&3.13&0.092&857.1$\pm$47.8&30051-01-02-00&30051-01-03-00&05-21 12:39&05-24 19:21&29\\
   &&&&&30051-01-03-01&30051-01-03-02&&&\\
3&-&-&-&-&30051-01-04-00&30051-01-04-01&05-28 08:33&05-28 14:30&2\\
4&0.54&3.10&0.085&864.2$\pm$53.7&30051-01-05-00& &06-01 09:60&06-01 17:40&28\\
5&0.60&3.27&0.249&824.4$\pm$36.2&30051-01-06-00&30051-01-07-00&06-05 05:16&06-09 12:42&25\\
   &&&&&30051-01-07-01&30051-01-07-02&&&\\
6&0.51&3.14&0.086&829.2$\pm$49.9&30051-01-08-00&30051-01-08-01&06-13 01:26&06-17 15:29&110\\
   &&&&&30050-01-01-00&30050-01-01-01&&&\\
   &&&&&30050-01-01-02&30050-01-01-03&&&\\
   &&&&&30050-01-01-04&30050-01-01-050&&&\\
7&0.51&3.14&0.086&837.3$\pm$50.7&30050-01-01-05&30050-01-01-06&06-17 15:58&06-20 10:19&69\\
   &&&&&30050-01-01-07&30050-01-01-080&&&\\
   &&&&&30050-01-01-08&30050-01-02-00&&&\\
   &&&&&30050-01-02-01&30050-01-02-02&&&\\
   &&&&&30050-01-02-03&&&&\\
8&0.46&3.08&0.010&756.0$\pm$50.0&30051-01-09-00&30051-01-09-01&06-26 09:34&06-29 16:09&41\\
   &&&&&30051-01-10-00& &&&\\
9&0.66&3.30&0.305&842.7$\pm$20.6&30051-01-11-00&30051-01-11-01&07-03 14:22&07-09 15:40&78\\
   &&&&&30051-01-11-02&30051-01-11-03&&&\\
   &&&&&30051-01-12-00& &&&\\
10a&-&-&-&-&30051-01-12-01&&07-13 09:36&07-13 15:39&22\\
10b&-&-&-&-&30051-01-12-02&&07-31 15:11&07-31 18:59&14\\
10c&-&-&-&-&30050-01-03-00&&08-09 13:31&08-09 17:25&14\\
11a&-&-&-&-&30050-01-03-01&&08-14 13:18&08-14 17:55&17\\
11b&-&-&-&-&30050-01-03-02&&08-21 13:02&08-21 16:11&11\\
12&0.49&3.14&0.076&829.7$\pm$33.6&30050-01-03-03& &10-10 01:45&10-10 06:44&18\\
\hline
\end{tabular}
\end{flushleft}
\caption[]{The RXTE observation IDs of the observations that are included in each track, the HC and SC of their vertices, and the distance, $S_\mathrm{shift}$, over which each track is shifted relative to track 1. Average flux at the vertex ($S_\mathrm{a}=2$), start time of the first and end time of the last included observations and the total included observation time are also indicated. Note that tracks 3, 10a-c and 11a-b do not show a vertex. All dates are in the year 1998; typical errors in HC are $\sim0.01$, in SC$\sim0.03$, in $S_\mathrm{shift}\sim0.01$ and in flux at $S_\mathrm{a}=2\sim43$.}
\label{ptracks}
\end{table*}

\FloatBarrier

\begin{table}
\begin{flushleft}
\begin{tabular}{cccccc}
\hline
\hline
Track & $S_\mathrm{a}$ & Sign. & $\nu_\mathrm{max}$ & Amplitude & $Q$\\
 & & ($\sigma$) & (Hz) & (\% rms) & \\
\hline
1&$1.38{\pm0.03}$&4.3&$66^{+5}_{-4}$&$2.3{\pm0.3}$&$2.0^{+0.8}_{-0.6}$\\
1&$1.47{\pm0.04}$&3.8&$66^{+5}_{-4}$&$1.9{\pm0.3}$&$2.2^{+1.4}_{-0.8}$\\
2&$1.51{\pm0.05}$&3.7&$69{\pm2}$&$1.8{\pm0.3}$&$4.5^{+2.3}_{-1.5}$\\
2&$1.64{\pm0.04}$&3.5&$73^{+4}_{-3}$&$1.5{\pm0.2}$&$3.7^{+2.7}_{-1.5}$\\
2&$1.76{\pm0.03}$&2.7&$74{\pm3}$&$1.3{\pm0.3}$&$5^{+6}_{-2}$\\
6&$1.41{\pm0.05}$&7.4&$66.2^{+1.2}_{-1.3}$&$2.7{\pm0.2}$&$4.0^{+0.9}_{-0.7}$\\
6&$1.60{\pm0.07}$&6.7&$65.5^{+1.6}_{-1.5}$&$2.05^{+0.15}_{-0.16}$&$3.7^{+0.9}_{-0.7}$\\
7&$1.31{\pm0.05}$&6.7&$63.9{\pm0.9}$&$2.20{\pm0.17}$&$4.4^{+1.0}_{-0.8}$\\
7&$1.50{\pm0.08}$&11.2&$64.2{\pm0.7}$&$2.37^{+0.10}_{-0.11}$&$3.5^{+0.5}_{-0.4}$\\
7&$1.76{\pm0.08}$&2.7&$64.3^{+5.7}_{-1.4}$&$1.0^{+0.4}_{-0.2}$&$9^{+36}_{-7}$\\
8&$1.16{\pm0.06}$&10.0&$64.1^{+1.1}_{-1.0}$&$2.75^{+0.13}_{-0.14}$&$3.0^{+0.4}_{-0.3}$\\
8&$1.38{\pm0.08}$&4.4&$62{\pm3}$&$2.2^{+0.2}_{-0.3}$&$2.6^{+1.2}_{-0.7}$\\
\hline
\end{tabular}
\end{flushleft}
\caption[]{All detected QPOs and their properties. Typical 95\% confidence upper limits ($\nu_\mathrm{max}=68$ Hz, $Q=3$) for tracks in which the QPO was not detected are up to $\sim$2\% rms.}
\label{qpo_prop}
\end{table}

\end{document}